\pdfoutput=1
\documentclass[acmlarge,nonacm]{acmart}
\usepackage{parskip}
\begin{document}
\title{Mind the Gap: Foundation Models and the Covert Proliferation of Military Intelligence, Surveillance, and Targeting}
\author{Heidy Khlaaf}
\email{heidy@ainowinstitute.org}
\affiliation{%
  \institution{AI Now Institute}
}
\author{Sarah Myers West}
\email{sarah@ainowinstitute.org}
\affiliation{%
  \institution{AI Now Institute}
}
\author{Meredith Whittaker}
\affiliation{%
  \institution{Signal}
}
\authorsaddresses{}
\renewcommand{\shortauthors}{Khlaaf, Myers West, and Whittaker}

\begin{abstract}
Discussions regarding the dual use of foundation models and the risks they pose have overwhelmingly focused on a narrow set of use cases and national security directives---in particular, how AI may enable the efficient construction of a class of systems referred to as CBRN: chemical, biological, radiological, and nuclear weapons. The focus on these hypothetical and narrow themes has occluded a much-needed conversation regarding present uses of AI for military systems, specifically ISTAR: intelligence, surveillance, target acquisition, and reconnaissance. These are the uses most grounded in actual deployments of AI that pose life-or-death stakes for civilians, where misuses and failures pose geopolitical consequences and military escalations. This is particularly underscored by novel proliferation risks specific to the widespread availability of commercial models and the lack of effective approaches that reliably prevent them from contributing to ISTAR capabilities.

In this paper, we outline the significant national security concerns emanating from current and envisioned uses of commercial foundation models outside of CBRN contexts, and critique the narrowing of the policy debate that has resulted from a CBRN focus (e.g., compute thresholds, model weight release). We demonstrate that the inability to prevent personally identifiable information from contributing to ISTAR capabilities within commercial foundation models may lead to the use and proliferation of military AI technologies by adversaries. We also show how the usage of foundation models within military settings inherently expands the attack vectors of military systems and the defense infrastructures they interface with. We conclude that in order to secure military systems and limit the proliferation of AI-based armaments, it may be necessary to insulate military AI systems and personal data from commercial foundation models.
\end{abstract}
\maketitle

\section{Introduction}
As with the vast majority of technologies, AI models have dual use in both civilian and military applications. In particular, commercial foundation models~\footnote{Foundation models are "general-purpose AI" models capable of a range of general tasks including text, image, or audio generation and manipulation. Examples include OpenAI’s ChatGPT and Dall-E.}, trained on both private and public data, are being repurposed for military uses~\cite{vincent_inside_2023}. Not surprisingly, defining governance interventions for said AI systems has become the subject of a highly polarized debate~\cite{hao_new_2023}. On one hand, some argue that regulatory intervention might effectively impede innovation~\cite{rstreet,mozilla_releasing}, while others support imposing the strictest possible measures on AI to prevent the realization of future catastrophic risks~\cite{sc_staff_bill_2024,perrigo_u._2024,ai_now_institute_tracking_2023}. Despite continuing exuberance for military AI across industry and government~\cite{startup_nodate,MoD_2022,dod_2023}, conversations about the development of autonomous systems or AI armaments have neglected to apply claim-oriented approaches~\footnote{Approaches that assess the fitness of a given system against a scoped and falsifiable claim traditionally deployed by safety-critical and defense domains.} to substantiate assertions concerning the fitness of AI systems within military contexts~\cite{khlaaf_safety_2024}, and the efficacy of respective policies aimed at AI nonproliferation and safeguarding national security.

Yet, this politicized debate has largely evaded fundamental subjects such as the consideration of risks beyond chemical, biological, radiological, and nuclear weapons (CBRN) that may compromise national security. This has in turn led to a narrowing of the policy debate that over-indexes on several measures that have predominantly relied on the application of quantified compute thresholds---which falsely correlate compute quantities to the capabilities of AI models---or on limiting the public release of model weights. However, these approaches lack basic clarity on the fundamental barriers to the efficacy and administrability of policy interventions, and do not prevent sensitive and dual-use data, such as personally identifiable information~\footnote{While the term PII has varied definitions in the US regulatory context, in this paper we use it to broadly refer to information that can be used to distinguish or trace a person's identity. See, Dept of Defence, Privacy, Civil Liberties, and FOIA Directorate.}, from contributing to the proliferation of AI-based intelligence, surveillance, target acquisition, and reconnaissance (ISTAR) capabilities for military operations~\cite{NATO_joint_2024}.

Recent examples of these ISTAR systems include Gospel, Lavender, and Where’s Daddy~\cite{abraham_lavender:_2024}, which have used AI to facilitate a significant civilian death toll in Gaza through the fallible collection and use of personally identifiable information. While Gospel, Lavender, and Where’s Daddy are not foundation models themselves, they have provided a precedent for error-prone AI predictions that result in high civilian casualties~\cite{gaza_algorithms_2024}. This precedent now persists through to commercial foundation models, which are being proposed to “help Pentagon computers better ‘see’ conditions on the battlefield, a particular boon for finding---and annihilating---targets”~\cite{biddle_microsoft_2024,poulson_leak_2024}. Such uses are illustrative of the expanding testing grounds for the increasing use of foundation models within ISTAR contexts that may amount to further civilian harm. Moreover, \textbf{the accessibility of foundation models entails that the lack of consideration of personal information in the same vein as chemical, biological, radiological, or nuclear data within them neglects a crucial proliferation risk vector previously unseen in other AI technologies. Subsequent governance interventions are thus required to prevent the proliferation and unfettered scaling of ISTAR misuses and failures stemming from foundation models that may result in deadly and geopolitically consequential impacts, and may bolster potential military escalations.}

In this paper, we provide a claim-oriented analysis that the fixation on hypothetical CBRN weapons~\cite{irving_red-teaming_2024} has not only narrowed the scope for proliferation interventions in a manner that has led to an over-indexing on several measures, but has also occluded how the risk of personal data being embedded within existing commercial foundation models positions AI as a link between commercial personal data and automated weapons’ target lists and surveillance capabilities. This repurposing of dual-use commercial foundation models in military contexts, the current use of personal information in the training of large commercial foundation models, and the inability or unwillingness to protect or excise personal information from training datasets~\cite{lukas_analyzing_2023}, renders controls designed to prevent AI proliferation infeasible. Additionally, we provide an evidence-based case that usage of commercial foundation models inherently expands the attack vectors adversaries can use to exploit safety-critical systems, including AI military systems and the defense infrastructures they interface with. 

We conclude with key considerations for national security policy in AI. That is, the aperture of policy analysis must reprioritize to consider the use and proliferation of AI systems in ISTAR contexts leveraged through foundation model capabilities, and the risks and attack vectors that this creates. As such, for AI nonproliferation controls to be effective and to reduce national security risks, policymakers must consider the elimination of personal data from within the training data used to create foundation models, and that it may be necessary to insulate military AI systems and personally identifiable information from commercial foundation models.
\section{Background: AI within a longer history of governing interventions}

Debates relating to governance interventions supporting the US’s national security and foreign policy through the control of software and hardware technologies are not novel, and frequently feed into “arms race” narratives around the proliferation of technological capability and expertise~\cite{ai_now_institute_tracking_2023}. In this section, we provide a brief history of said interventions in the context of traditional technologies (e.g., hardware and software) and their relation to AI technologies, followed by recently proposed AI-tailored interventions that characterize the ineffective narratives we address in the remainder of the paper.

\subsection{Hardware and Software Export Controls}
Unilateral export controls have often been used to regulate the export or transfer of commercial and military items in order to limit the proliferation of weapons of mass destruction and destabilizing accumulations of conventional weapons and dual-use technologies~\cite{department_of_state_information_overview_2011}. In the 1970s, the US enacted various export control policies on high-performance computing (HPC) aimed at the Soviet Union~\cite{daniels_safeguarding_2022}. In the 1990s, a decades-long debate about the role of encryption, and the tight export controls then governing it, ultimately resulted in the liberalization of strong encryption~\cite{doomed}. Most recently, the US instituted a set of restrictions aiming to limit the proliferation of leading-edge semiconductors, alongside the passage of legislation allocating federal funding toward homegrown manufacturing~\cite{kak_modern_2024}. Additionally, certain types of semiconductors used in the training of AI fall under the United States’ export control regime~\cite{export_2023}. More generally, foreign-made items incorporating 25 percent or more of controlled US origin content are potentially subject to the Export Administration Regulations (EAR) for purposes of export or reexport of items falling under the Commerce Control List (CCL)~\cite{bis_nodate}. Similar controls exist under the US’s International Traffic in Arms Regulations (ITAR) that specifically address surveillance, weapons, or defense articles. Suffice it to say that myriad restrictions and caveats governing the sale and purchase of computational technology currently exist, shaping markets and processes across the industry.

The proliferation of AI poses many of the same challenges historically observed with other technological systems, in addition to unique considerations that must be accounted for in the process of defining effective export controls. \textbf{Contrary to popular belief, AI-based systems do not possess any unique or distinct software or hardware elements that could warrant restrictions on the use of AI subcomponents without impeding the use of traditional software and hardware components as a whole}. In fact, deep neural networks (DNNs), which serve as the architectural basis of current foundation models, were initially developed between the 1960s  and the 1980s~\cite{ivachnenko_cybernetics_1967,amari_theory_1967,rumelhart_learning_1986}, while recent advancements in graphics processing units (GPUs) have enabled the capabilities of foundation models present today. In other words, AI is comprised of familiar subsystems that have established processes for reviewing and assessing, and existing controls already cover many of the subcomponents that make up current AI systems. 

What is unique to AI systems is \textit{what they learn} (i.e., data), rather than \textit{how they learn} (e.g., software and hardware components). Although existing controls may be sufficient for the latter (i.e., the subset of AI subcomponents), the issues we must confront lie within the training data that provides the underlying specifications, and subsequently the risks, that shape the behavior of AI systems. A focus on training data is key to the efficacy of any proposed export controls and necessitates keen attention to the AI supply chain, including the human labor tasked with making critical determinations in all operations of the model and data pipeline.

\subsection{Recent AI Governing Interventions}
Despite this, recent policy interventions with unclear efficacy, such as the US’s Executive Order on the Safe, Secure, and Trustworthy Development and Use of Artificial Intelligence, focus on self-reporting requirements for compute transactions and dual-use models that utilize “a quantity of computing power greater than 10\textsuperscript{26} integer or floating-point operations” or 10\textsuperscript{23} integer or floating-point operations for those “using primarily biological sequence data”~\cite{the_white_house_executive_2023}. These compute thresholds have often been proposed as regulatory pathways for foundation models by requiring licensing to access a large amount of compute for specific AI purposes~\cite{anderljung_frontier_2023}.

Interventions that specifically address data, such as the US’s Executive Order on Preventing Access to Americans’ Bulk Sensitive Personal Data, outline explicit restrictions on the licensing, transfer, and export of bulk sensitive personal data or US Government-related data~\cite{the_white_house_executive_2024}. This executive order ultimately seeks to restrict specific types of data transactions between US persons and “countries of concern” that “can also use access to bulk data sets to fuel the creation and \textit{refinement of AI} and other advanced technologies, thereby improving their ability to exploit the underlying data and exacerbating the national security and foreign policy threats”. However, these restrictions do not extend to AI models themselves, despite primarily being a representation of their training data and the inherent vulnerabilities within foundation models that allow for the extraction of model data through observed model predictions alone~\cite{carlini_extracting_2023}. 

In the following sections, we survey key limitations on how proposed AI nonproliferation interventions have not only occluded a much-needed conversation regarding the efficacy of proposed measures, but also disregard a larger class of dual uses of commercial foundation models in military contexts, namely ISTAR operations.

\section{Ramifications of CBRN over-indexing and its interventions}

Current policy regarding the risk of AI-based military proliferation has focused
on the quantification of compute thresholds---which falsely correlate compute quantities to the capabilities of AI models---and on potentially limiting the public release of model weights. These concerns, and the subsequent proposed governance interventions, have largely stemmed from hypothetical CBRN risks regarding the ability of foundation models to construct or manufacture chemical, nuclear, or biological weapons. Yet these approaches lack basic clarity on the fundamental barriers to the efficacy and administrability of such interventions, and do not prevent sensitive data from contributing toward the proliferation of intelligence, surveillance, target acquisition, and reconnaissance capabilities.

The underlying presumption of these interventions largely hinges on the unproven hypothesis that with increased levels of compute, unforeseen new AI capabilities (and thus risks) are introduced~\cite{irving_red-teaming_2024}. Specifically, it is speculated that sufficiently powerful foundation models may enable the proliferation of AI-based military capabilities through lowering the barrier of entry to design, synthesize, acquire, or use CBRN weapons~\cite{heim_accessing_2023,sastry_computing_2024}. As a result, the aforementioned US Executive Order on the Safe, Secure, and Trustworthy Development and Use of Artificial Intelligence outlines self-reporting requirements for models that utilize “a quantity of computing power greater than 10\textsuperscript{26} integer or floating-point operations” or 10\textsuperscript{23} operations for models using biological sequence data. The executive order further emphasizes that for national defense and the protection of critical infrastructure, oversight is required for “the ownership and possession of the model weights of any dual-use foundation models, and the physical and cybersecurity measures taken to protect those model weights”.

\textbf{Despite the heightened concern around CBRN weapons, uses of narrow compute thresholds to determine regulatory scope effectively exempt current military use of foundation models from the reporting and disclosure mandates of the executive order}. In fact, almost no existing models pass this compute threshold~\cite{sastry_computing_2024}. Claims that these thresholds are sufficient for prospective foundation models that may develop CBRN capabilities are discounted by demonstrations that equally capable replicas of existing large AI models can be developed without approaching this compute requirement, and that compute thresholds do not necessarily translate to capabilities~\cite{papernot_sok:_2018,hooker_limitations_2024}. Compute-based measures thus cannot act as a meaningful threshold intended to benchmark foundation model risks, including the emergence of and access to military capabilities, that would trigger consequential changes to US licensing requirements or export controls.

Policy interventions positing that limiting the public release of model weights protects national security interests also remain unsubstantiated~\cite{ntia_dual-use}. Momentarily putting aside the constitutive lack of accuracy and known failure modes of foundation models~\cite{weidinger_ethical_2021,khlaaf_hazard_2022}, they are additionally vulnerable to a wide array of attacks that render discussions on model weight publication futile. Adversaries are currently capable of carrying out membership attacks~\footnote{A black-box attack where an adversary probes whether or not a specific point was part of the training dataset analyzed to learn the model’s parameters.}, model inversion~\footnote{Also known as a training data extraction attack, where adversaries are able to extract training data from model predictions.}, and model extraction~\footnote{Where an adversary uses a model’s outputs as training data into their own model, that allows them to replicate the original model’s behavior.} attacks to extract training data and model parameters. Such attacks do not require access to model weights or data, and can be performed using observed model predictions or API access alone~\cite{carlini_stealing_2024,carlini_extracting_2023,papernot_sok:_2018,qi_fine-tuning_2023}. This would hinder the efficacy of any interventions that rely on restricting the export or transfer of model weights and data, given that the commercial foundation models used to underpin military-tuned models are publicly available and accessible, whether those models are open or closed source~\cite{gray_widder_open_2023}.

Finally, the aforementioned policy interventions neither account for nor apply to the construction of narrowly tailored models built for specific military purposes that are more effective at a given task~\cite{agarwal_ieee,Zhou_Schellaert_2024}. Deep neural networks (DNNs), which form the basis of current foundation models, cannot solve tasks outside of their data distribution and training data sets. Commercial AI labs are thus collaborating with states and militaries to fine-tune commercial foundation models using military data in pursuit of adapting the capabilities of “general” models for military or battlefield operations~\cite{vincent_inside_2023,biddle_microsoft_2024}. Furthermore, emergent military capabilities may arise out of data already embedded within existing foundation models, including personally identifiable information that may contribute toward the AI proliferation of ISTAR capabilities vital for military operations. 

This disregard for a larger class of dual uses of commercial foundation models has led to an oversight of crucial national security risk vectors, and subsequent governance interventions, required to prevent the proliferation of AI-based ISTAR misuses and failures that may result in increased military escalations and geopolitically consequential impacts.

\section{Personal Data and the Proliferation of AI ISTAR Armaments}

ISTAR is a class of military applications that aims to provide military decision-makers with situational awareness of the conditions on the ground, in the air, at sea, in space, and in the cyber domain~\cite{NATO_joint_2024}. ISTAR systems assist a military force in employing its sensors and managing the information they gather to link battlefield functions and operations. Examples of such systems include operational aircrafts like the MQ-9 Reaper, which is a remotely piloted medium-altitude, long endurance (MALE) aircraft designed for intelligence, surveillance, target acquisition and reconnaissance, and attack missions. Despite ISTAR being the use case most grounded in actual deployments of AI armaments, the proliferation of AI-based ISTAR systems by means of the availability and use of foundation models has largely been absent from governance and nonproliferation conversations. Yet, these are the uses that currently pose life-or-death stakes for civilians around the world, where AI misuses and failures will exacerbate deadly and geopolitically consequential impacts and bolster potential military escalations.

Whether the exclusion of ISTAR is simply a means to capitulate regulatory efforts or merely a lack of consideration of a larger class of dual uses of foundation models, the futility of limiting compute or hiding weights to guard against model capabilities necessitates a refocus on data provenance techniques as an alternative approach to nonproliferation for all classes of AI armaments~\cite{gebru_datasheets_2021}. The significance of data extends beyond its centrality to both specifying and understanding AI models, as its relevance has always been fundamental to general military technologies and nonproliferation efforts, AI-enabled or otherwise. As such, information and data are included within the items and software subject to US export control laws under ITAR and EAR. Indeed, military-relevant data for some AI armaments may already be covered by existing munitions lists (i.e., 22 CFR 121.1---the United States Munitions List, or USML). 

One approach might be to broaden the existing constraints on providing, sharing, or selling of data relevant to AI-enabled military capabilities to prevent the proliferation of AI weapons. These could include introducing explicit restrictions on data used in AI systems that support the following ISTAR military applications:
\begin{itemize}
\item Intelligence
\item Surveillance and Reconnaissance
\item Multi-domain Command and Control
\item Communications and Computers
\item Sub-threshold Information Advantage
\item Access and Maneuver
\end{itemize}
However, a particular---and likely existential---challenge arises in applying these interventions to commercial foundation models. Commercial foundation models’ training data often consists of personal information and activities collected from civilians, whether publicly scraped or procured through data brokers~\cite{ayoub_closing_2024,lukas_analyzing_2023,noauthor_algorithms_2024}. Personal data collected for creating targeted ads, tuning content algorithms, and serving search queries and the like enable dual-use capabilities that extend beyond commercial use. That is, personal data not only serves commercial ends but also serves as valuable intelligence and surveillance information that can be utilized for AI-enabled ISTAR systems.

\textbf{Personal data embedded within existing commercial foundation models thus positions AI as a link between commercial personal data and automated weapons’ target lists and surveillance capabilities}. These emergent military capabilities are already at play for US military uses, where it has been proposed that “images conjured by DALL-E could help Pentagon computers better ‘see’ conditions on the battlefield, a particular boon for finding — and annihilating — targets”
~\cite{biddle_microsoft_2024}. Other proposed use cases include the defense contractor Primer Technologies advertising its “next-gen AI” platform to automatically generate targeting reports after ingesting Open Source Intelligence (OSINT), a public source notorious for its inaccuracies and misinformation~\cite{primer_introducing_2023,poulson_leak_2024,armyosint}. And most recently, Scale AI has presented to the United States Special Operations Command an “AI ammo-factory” to support AI and autonomy missions across all programs within the US Department of Defense~\cite{poulson_leak_2024}. 

Additionally, the commercial availability of foundation models may enable adversaries to leverage these emerging targeting and surveillance capabilities to produce determinations and insights about populations whose data may have been trained on. This concern is further exacerbated by the previously aforementioned adversarial ability to carry out membership attacks, model inversion, and model extraction attacks to extract model training data. Although methodologies on machine unlearning have made attempts to construct an “unlearning” process by strategically limiting the influence of a data point in the training procedure~\cite{bourtoule_machine_2019}, recent research has demonstrated that these updates, instead, expose individual data points to high-accuracy reconstruction attacks that allow the attacker to recover this intimate data in its entirety~\cite{bertran_reconstruction_2024}. \textbf{Put bluntly, even with additional data restrictions in place, no effective approaches exist that reliably prevent personal data exposure in current foundation models, whether fine-tuned or otherwise, from contributing to ISTAR military capabilities}.

\subsection{Geopolitical Consequences of AI ISTAR Proliferation}
This inadequacy of available interventions inhibits the efficacy of oversight fundamental to preventing proliferation where, in this particular case, AI misuses and failures can exacerbate deadly and geopolitically consequential impacts and bolster potential military escalations. Indeed, such phenomena have already been observed with the use of Gospel, Lavender, and Where’s Daddy, which have been used to facilitate a significant civilian death toll in Gaza~\cite{abraham_lavender:_2024}. Gospel, Lavender, and Where’s Daddy are AI ISTAR military decision-making systems that have been touted as vital for the IDF’s military operations, all of which hinge on the use of personally identifiable information. Although not foundation models themselves, these systems have provided a geopolitical precedent where relying on personal data to produce error-prone AI predictions can be deployed to disastrous ends. In fact, these systems’ reliance on faulty data and “inexact approximations to inform military actions” could contravene Israel’s obligations under international humanitarian law~\cite{gaza_algorithms_2024}.

This precedent has already persisted through to foundation models, with unique proliferation consequences given the availability of commercial models and the lack of effective approaches that reliably prevent personal data from contributing to ISTAR capabilities. As previously noted, models are being used to automatically generate targeting reports after ingesting OSINT~\cite{primer_introducing_2023,poulson_leak_2024}, despite event barraging, misinformation campaigns, trend hijacking, and military deception fundamentally detracting from the usefulness of OSINT at the tactical level~\cite{armyosint}. Other promoted uses of foundation models within the ISTAR ecosystem include analyzing “real-time open-source data streams, and pinpoint potential escalation areas” within conflicts~\cite{laufer_bridging_2023}. However, studies have already demonstrated that even within simulations, off-the-shelf foundation models often produce decisions that encourage conflict escalation and escalation patterns that are difficult-to-predict~\cite{rivera_escalation_2024}. These interactions realized on the battlefield may cause unintentional military escalations that have adverse consequences, especially in the case of a face-off between nuclear-armed states~\cite{boulanin_impact_2019}. 

Finally, commercial foundation models can be weaponized against state citizens whose data these models have been trained on~\cite{nasr_scalable_2023}. Marginalized people are particularly imperiled, as their status inherently puts them at risk of being targeted and surveilled by adversaries who may use commercial models to produce determinations and insights about them. Realizations of these uses are no longer hypothetical and have recently come to light with the use of commercially available facial image recognition services to derive URLs that feed into a foundation model “to infer the person’s name, job, and other personal details”~\cite{cox_someone_2024}. As such, these proliferation risks necessitate the establishment of novel interventions that specifically address the repurposing of dual use commercial foundation models and their data within military contexts.

\subsection{Toward Effective ISTAR Proliferation Interventions}
Given the stakes here, both in terms of proliferation and human cost, continuing to train commercial foundation models on sensitive personal data and civilian activities will only bolster and leverage the AI capabilities of external adversaries (e.g., espionage, surveillance). As Craig Martell from the US Department of Defense Digital and Artificial Intelligence Office noted, “any input into publicly accessible Gen AI tools is analogous to a public release of that information”~\cite{poulson_leak_2024}. It is thus in the interest of states and militaries worldwide to consider further restrictions on the use, dissemination, and export of both data and models---whether through export controls or other legislative measures---to prevent their citizens’ personal information from being utilized for AI-enabled military capabilities. Existing US interventions, such as the US’s Executive Order on Preventing Access to Americans’ Bulk Sensitive Personal Data~\cite{the_white_house_executive_2024}, have already considered explicit restrictions on the transfer and export of personal data, and acknowledge the role of personally identifiable information in the creation and refinement of AI given that “countries of concern can rely on … AI, to analyze and manipulate bulk sensitive personal data to engage in espionage, influence, kinetic, or cyber operations or to identify other potential strategic advantages over the United States.” However, these restrictions do not extend to AI models themselves, despite foundation models embedding representations of their training data and, subsequently, potential military capabilities within their functionality. Furthermore, this framework fails to recognize the inherent vulnerabilities within foundation models that allow for the extraction of model data through observed model predictions alone~\cite{carlini_extracting_2023}.

In developing guidelines for the appropriate use of AI in national security contexts, policymakers will need to consider interventions that will prevent personal data within foundation models from contributing to military capabilities to ensure effective and administrable controls over these technologies. In doing so, they will encounter potentially existential limitations to the usage, procurement, and regulation of commercial foundation models in military contexts: \textbf{it may be impossible to guarantee the security of these systems to the level of assurance needed for military deployment, and their implementation may in fact introduce greater risks to the infrastructures in which they are embedded that are disproportionate to the benefits that any use of AI may produce}.
\section{The Expansion of National Security Attack Vectors with Dual-Use Models}

In developing the aforementioned guidelines for the appropriate use of AI in national security contexts and respective administrability controls over these technologies, policymakers must consider the inherent vulnerabilities of commercial foundation models that expand attack vectors that adversaries can use to exploit AI military systems and the defense infrastructure they interface with. Such expanded vectors of attacks include theoretical and practical demonstrations of “jailbreaks” and adversarial attacks that aim to craft inputs that manipulate a model to produce intentionally erroneous outputs or subvert its safety filters and restrictions~\cite{el-mhamdi_impossible_2022,wu_attacks_2024}. Other new and undetectable attack vectors include poisoning web-scale training datasets and “sleeper agents” within commercial foundation models, which may intentionally or inadvertently assist the subversion of models used within military applications and ultimately compromise their behavior~\cite{noauthor_sleeper_agents,carlini_poisoning_2024}. 

Several approaches have attempted to address these challenges to no avail~\cite{ganguli_red_2022}, as research has persistently shown that it is always possible to construct attacks that are transferable across all existing foundation models~\cite{zou_universal_2023}. As a result, any fine-tuning or guardrails introduced as a way to enable accurate military performance or security protections could be bypassed. Potential existential limitations in combating these novel attack vectors also arise due to the lack of traceability of human labor and unknown data sources across the supply chain of commercial foundation models repurposed for AI military applications. Indeed, traceability~\footnote{Traceability is the procedure of tracking and documenting all artifacts throughout development and manufacturing processes.}, a core requirement of military and safety-critical systems, is required to guarantee that no aspect of the development pipeline is compromised to ensure a system’s security and fitness for use.

Consider the significant use of human labor and involvement across the AI development cycle. Discussions of both Human-In-The-Loop (HITL)~\cite{crootof_humans_2022}, in line with the definition regulators often implicitly employ, or Meaningful Human Control (MHC) for autonomous weapons systems~\cite{amoroso_toward_2021} recognize the involvement of individuals who oversee particular decisions made in conjunction with an algorithm. Yet absent from these discussions is how those creating, curating, or fine-tuning data and building infrastructure within the AI supply chain may be adversarially leveraged to introduce vulnerabilities and backdoor attacks such as data ordering and model-spinning attacks~\cite{shumailov_manipulating_2024}, or by purposely or inadvertently providing malicious or poor scoring and feedback in Reinforcement Learning from Human Feedback (RLHF)~\cite{ouyang_training_2022} to degrade or selectively target model performance. More generally, the ubiquitous and unfettered use of web-scale datasets for training commercial foundation models has led to the exploitation and use of several avenues that allow adversarial actors to execute poisoning attacks “that guarantee malicious examples will appear in web-scale datasets”~\cite{carlini_poisoning_2024}. 

The myriad identified vulnerabilities and backdoor attacks are symptomatic of how commercial foundation models lack the appropriate traceability between model development and the respective data and labor used from their source and point of manufacturing (including RLHF). As such, the repurposing of dual-use commercial foundation models in military contexts would inhibit not only the operationalization of AI interventions seeking to control and halt the proliferation of AI-armaments, but also the security guarantees typically required for military systems. This lack of traceability of commercial foundation models emphasizes the potential need to separate models for commercial use from those used for military applications in order to uphold national security. For example, the Navy’s chief information officer, Jane Overslaugh Rathbun, has noted that commercial models have “inherent security vulnerabilities” and are “not recommended for operational use cases”~\cite{navy_department}.

Given this lack of appropriate traceability of commercial foundation models, states may choose to build their own military-exclusive (i.e., non-commercial) foundation models for ISTAR purposes. However, military-exclusive models would not remediate against the inaccuracies of DNNs and the reality of attacks that extract model data through observed model predictions alone~\cite{carlini_extracting_2023}. As such, stakeholders not only need to consider the threat models of their military use cases and whether an adversary could utilize observations of model outputs in pursuit of AI armaments, but should also look to alternative policy paradigms to reduce national security risks. We put forward such interventions in the next section.

\section{Key Considerations for National Security and AI}

Overall, the fixation on hypothetical CBRN weapons has not only narrowed the efficacy and scope of proliferation intervention~\cite{irving_red-teaming_2024}, but also has occluded the risks of how personally identifiable information within existing commercial foundation models can accelerate ISTAR-driven coercive automated decisions. In developing appropriate proliferation interventions that would be effective across a larger set of AI military use cases, policymakers will need to consider interventions that will prevent personal data within foundation models from contributing to military capabilities for effective and administrable controls over these technologies. However, in doing so, they will encounter existential limitations to the usage and regulation of commercial foundation models in military contexts given the expanded attack vectors that adversaries can use to exploit AI-based military systems and compromise national security.

This only accentuates the need for policy aimed at AI-based proliferation to prioritize the following considerations:
\begin{enumerate}
    \item Addressing the inclusion of personal information in training data for foundation models as a source of national security risk. Personally identifiable information enables models to be used for military capabilities such as ISTAR, as such information acts as valuable intelligence that may be utilized by adversaries to develop AI-enabled military systems that surveil and target specific populations. Existing precedent for data-protection rules may provide a basis to address usage of personal information within commercial AI models~\cite{iapp_2024,ftc_amazon}.
    \item Maintaining traceability---an existing core requirement of military and safety-critical systems---with any AI use to guarantee that no aspect of the development pipeline is compromised, and to ensure a system’s security and fitness for use. As such, implementing a traceability mandate for any AI-based military systems or armaments within national security contexts will be necessary to ensure robust security practices. Ensuring  the use of established security methodologies, such as secure development pipelines that can mitigate against the vulnerabilities identified thus far~\cite{brundage_toward_2020,milanov_exploiting_2024,sorensen_leftoverlocals:_2024}, is particularly important.
    \item Assessing whether military-exclusive foundation models can be developed without building on commercial sources, whether such models enable more accurate determinations, and whether they offer safeguarding of sensitive intelligence. This is necessary due to the inability to adequately trace the data and labor used in commercial foundation model development, opening models to vulnerabilities and backdoor attacks~\cite{carlini_poisoning_2024}.
    \item Constraining the flow of sensitive and personal data from citizens that can enable them to be targeted by adversaries using AI systems is in the national interest. To be effective, such constraints may also need to extend to models trained on this data. The US’s Executive Order on Preventing Access to Americans’ Bulk Sensitive Personal Data, has already considered explicit restrictions on the transfer and export of personal data~\cite{the_white_house_executive_2024}. However, commercially available models trained on such data can enable the insights derived from personal data to be utilized for military capabilities (e.g., targeting) even where the data itself is restricted from sale. To be effective, restrictions placed on personal data flows would also need to be extended to models trained on that data.
    \item Using foundation models in national security contexts may introduce unique concerns threatening human rights. For example, a government’s ability to train models on citizens’ data obtained through commercial data brokers that would otherwise need a warrant, court order, or subpoena to obtain may allow governments to further exercise coercive powers that are automated through AI decision-making~\cite{ayoub_closing_2024}. Such use may subvert due process, exacerbated when inaccurate outputs inflict unjust harms on civilians. Appropriate interventions may include the extension of data minimization principles to include purpose limitations on the collection, processing, and transfer of personal data to third parties for intelligence purposes~\cite{ayoub_closing_2024}.
    
\end{enumerate}

\bibliographystyle{ACM-Reference-Format}
\clearpage
\bibliography{citations}

\end{document}